# Can Recombination Displace Dominant Scientific Ideas?

Linzhuo Li[*], Yiling Lin[**], and Lingfei Wu[**]

[*] *newllqllz@gmail.com*
Department of Sociology, Zhejiang University, China

[**] *liw105@pitt.edu*
School of Computing and Information, University of Pittsburgh, United States

**Abstract.**

The combination of diverse, pre-existing knowledge is a common explanation for scientific breakthroughs. However, a paradox exists: while scientific output and the potential for such recombination have grown exponentially, the rate of breakthrough discoveries has not. To explore this paradox, our study examines 41 million scientific articles from 1965 to 2024. We measure two key properties for each paper: atypicality, which quantifies the combination of knowledge from conceptually distant areas, and disruption. We demonstrate that these metrics capture distinct processes. Atypicality is characteristic of work that extends established concepts into new topical areas (a form of cross-topic recombination). Disruption, in contrast, signifies the replacement of a dominant idea within its own topic.

## 1. Introduction

Understanding the origins of breakthrough ideas and their method of reshaping scientific knowledge—whether through extension or replacement of existing paradigms—is a central goal of innovation studies. This debate is largely framed by two opposing viewpoints: the recombination view and the displacement view.

The recombination view, formalized in recombinant growth theory, posits that innovation is an additive phenomenon driven by new combinations of existing knowledge. A significant limitation of this perspective is its omission of competitive idea dynamics. It struggles to differentiate between "normal science," which consolidates established concepts, and "scientific revolutions," which challenge them. The latter, not the former, is the hallmark of a breakthrough. The modern scientific landscape, characterized by surging publication volumes but slowing breakthrough rates, may exemplify this distinction. Explanations for this "productivity-progress" paradox, such as the "burden of knowledge" or misaligned funding incentives, have been proposed, but they often remain within the same additive framework, treating stagnation as a failure of broad search.

Conversely, the displacement view argues that the purpose of the search matters more than its scope. Progress can emerge from focused, local investigation when that investigation is aimed at challenging established knowledge. Foundational theorists of science have long argued that scientific advancement is not purely cumulative; new ideas frequently gain traction by rendering older ones obsolete. Einstein's general relativity, for instance, did not merely add to Newtonian mechanics; it supplanted it by resolving anomalies (like Mercury's orbit) that the previous framework could not. Under this framework, displacement, not recombination, is the engine of progress. However, this perspective has historically lacked scalable empirical operationalization, limiting its use in quantitative science studies.

The development of the Disruption Index (D-index) offers a solution to this measurement problem. The D-index is designed to quantify the degree to which a new paper diverts scholarly attention from the prior work it cites, effectively measuring if it serves as a new foundation for future research. It can, for example, distinguish between Watson and Crick's 1953 paper on DNA's double-helix structure (which was highly disruptive, $D = 0.96$) and the 2001 draft of the human genome (which was highly consolidating, $D = -0.005$). The former fundamentally altered its field (epistemic change), while the latter built upon and extended existing knowledge (intellectual succession). The D-index thus provides a quantitative tool for testing the displacement view.

We validate this displacement mechanism at scale using our full bibliometric dataset. By employing an LLM to calculate topical similarity, we find that highly disruptive papers show a strong topical correspondence with their most-cited reference, even after controlling for field, citation counts, and time. This supports the concept of targeted, within-domain replacement. Conversely, highly atypical papers are characterized by low topical similarity to their dominant references, reflecting a strategy of applying established ideas in new and distant contexts. This confirms that these two innovation modes are structurally distinct.

Finally, we use LLMs to classify papers as "theory" or "method" to explore the dynamics of idea displacement. In line with recent findings, our analysis shows that methodological papers are, on average, significantly more disruptive than theoretical papers. However, when we analyze the persistence of dominant ideas over time, we find that canonical theories are "harder to displace"—they remain dominant for longer periods than canonical methods. For example, the displacement of Gödel's (1931) theory of computation by Turing's (1937) occurred relatively quickly, while the displacement of Milgram's (1967) small-world experiment by Watts and Strogatz's (1998) graph model took over three decades. This highlights the foundational role of methodological advances, complementing Kuhn's focus on theoretical paradigm shifts.

In conclusion, our research argues that scientific progress relies on idea displacement, or selective forgetting, as much as it does on knowledge accumulation. While recombination is a frequent source of idea generation, our results challenge the assumption that broader recombination necessarily leads to more impactful innovation. Instead, we provide a new, empirically-grounded framework for understanding how new ideas emerge, gain legitimacy, and ultimately supplant their predecessors.

## Can Recombination Displace Dominant Scientific Ideas?

The most significant scientific breakthroughs often result not from simple refinement but from the replacement of a dominant idea with a functionally equivalent alternative. The cases of Turing (1937) and Barabási and Albert (1999) are illustrative. Turing's mechanical model of computation (the "Turing machine") supplanted Gödel's (1931) symbolic model. Likewise, the Barabási-Albert (1999) preferential attachment model replaced the Erdős–Rényi random graph model because it better captured the scale-free properties of real-world networks. In both instances, a new idea addressed the same core problem as its predecessor, but in a more tractable or compelling way.

This pattern highlights a central principle we call **functional equivalence**: displacement happens when a new concept can perform the same intellectual or epistemic job as a dominant predecessor, but more effectively. Recombination, from this standpoint, is an unlikely mechanism for displacement. Recombination typically joins complementary, conceptually distant sources. Displacement, however, requires a *substitutive* relationship. A novel combination (like a hypothetical "microwave TV") may be creative, but it is unlikely to *replace* either the microwave or the television. Displacement demands a new solution that competes directly with an old one for the same intellectual role.

We therefore predict that combinatorial novelty and displacement are structurally distinct and likely to be found in different types of scientific work. This leads to our first hypothesis:

**H1: Atypicality and Disruption Are Negatively Correlated.** We expect that papers drawing on distant, varied knowledge (high A-index) will be less likely to supplant dominant ideas (low D-index). This would confirm that recombination and displacement are separate, non-overlapping innovation pathways.

We then investigate the temporal dimension of epistemic replacement. Kuhn noted that while paradigm shifts are theory-centric, the anomalies that trigger them often arise from methodological, instrumental, or measurement advances. This implies a structural asymmetry: new tools may be required to identify the failures that catalyze theoretical change. If this is the case, theoretical frameworks may be more entrenched and persistent than the methods used to test them. This leads to our third hypothesis:

**H2: Theories are harder to displace than methods.** We predict that dominant theories, which are often embedded in broader conceptual paradigms, will demonstrate greater persistence (i.e., be displaced more slowly) than dominant methods, which are more vulnerable to replacement by new, more efficient techniques.

Taken together, these hypotheses provide a new framework for analyzing scientific change as a process of epistemic competition, contrasting the mechanisms of recombination and displacement.

## 3. Data and Methods

### Datasets

Our empirical strategy relies on four distinct and complementary data sources.

**Large-Scale Bibliometric Dataset.** The primary dataset is derived from the OpenAlex corpus with 49 million articles

### Measures

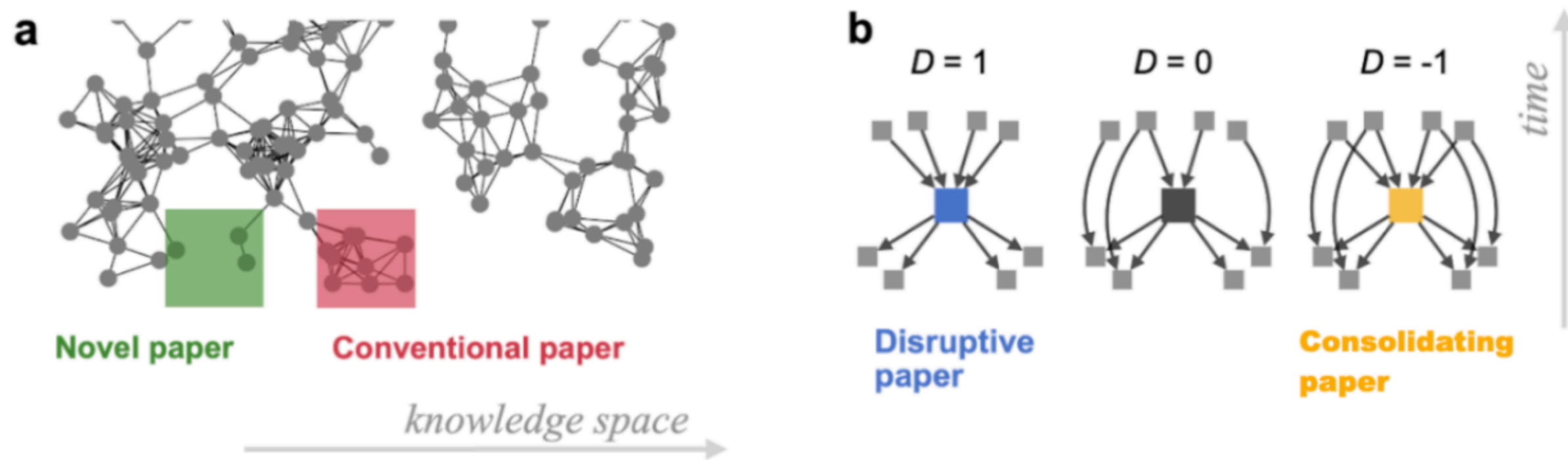

**Figure 1. Illustrations of Atypicality and Disruption. (a)** Atypicality reflects the extent to which a paper cites prior work from across the knowledge space, represented by co-citation networks. Novel papers (green) draw references from distant clusters, while conventional papers (red) cite within a single, localized cluster. **(b)** Disruption captures how a paper alters the trajectory of future knowledge, as represented in citation networks. Disruptive papers attract citations from future work that overlooks their own references, thereby eclipsing the influence of prior work. In contrast, developing papers are cited alongside their references, reinforcing the status of previous contributions.

**(1) Atypicality Index (A-index).** We use the Atypicality metric introduced by Uzzi et al. (2013), for each journal pair, we compute a z-score:

$$z_{ij} = \frac{obs_{ij} - exp_{ij}}{\sigma_{ij}}$$

**(2) Disruption Index (D-index).** We calculate disruption using the method originally proposed by Funk and Owen-Smith (2017) and later adapted and validated by Wu et al. (2019) to analyze scientific breakthroughs in paper citation graphs.

$$D_p = \frac{N_i - N_j}{N_i + N_j + N_k}$$

## 4. Results

### 4.1 The Rise of Atypicality and Decline of Disruption in Science Over Time

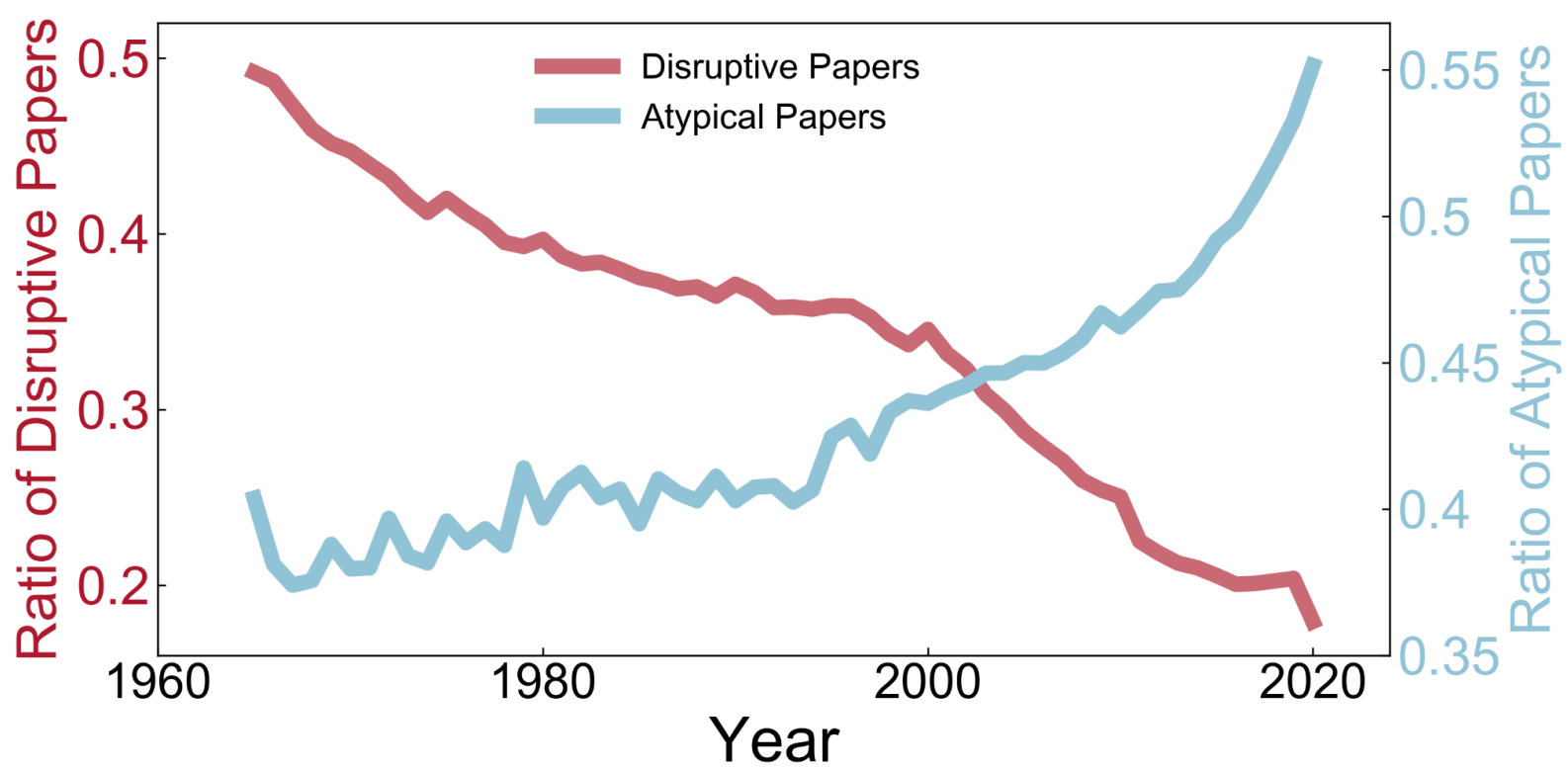

**Figure 2. The rise of highly atypical papers and decline of highly displacing Papers.** Analyzing 40,453,825 journal articles (1965-2020), we redraw the figure in the main text using the ratio of top 5% displacing papers (D > 0.04, in red) as well as top 5% atypical papers (A > 22.9, in blue) over time. The results are consistent with the figure in the main text.

## 4.2 Atypicality and Disruption Are Negatively Correlated

.

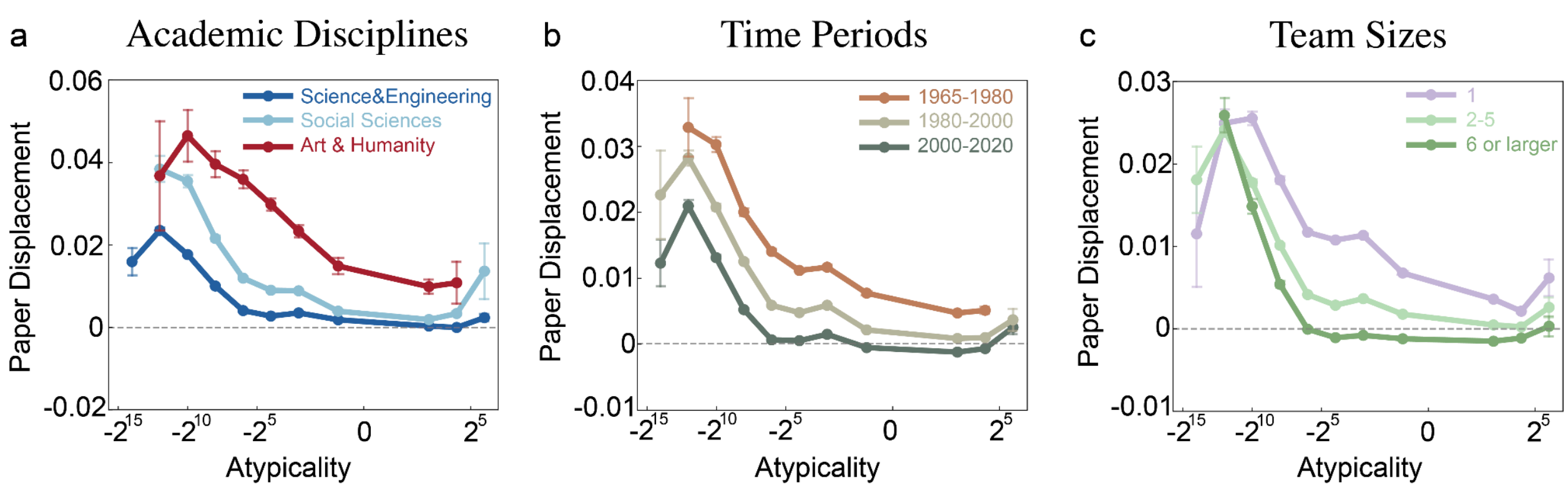

**Figure 3. Negative correlation between paper atypicality and disruption.** Analyzing 40,935,251 research articles from 1965 to 2020, we observe a consistent negative relationship between a paper's reference atypicality and its likelihood of displacing prior work. Each panel shows the average D-index within bins of A-index values, stratified by (a) academic discipline, (b) publication time period, and (c) team size. Across all groups, papers with highly atypical references tend to exhibit lower disruption scores. Error bars represent 95% bootstrap confidence intervals.

## 4.3 Breakthroughs from Theoretical and Methodological Displacements

To further explore the mechanisms underlying disruptive innovation, we examine whether conceptual and methodological advances follow different temporal patterns. For all 50,147 highly cited and disruptive (D > 0) papers in our dataset, we employed Llama-3.1-70B-Instruct, the state-of-the-art open-source large language model developed by Meta, to classify the type of breakthrough each paper represents relative to its most cited reference.

Using a zero-shot prompt, we provided the model with the title and abstract of each focal paper and its top reference, and asked:

> "Given two papers—Paper A {title, abstract} and its reference Paper B {title, abstract}—is Paper A better considered as (1) a 'theory' innovation (conceptually different from B), or (2) a 'method' innovation (a refinement or formalization of B)? Only give the option number."

We extracted the model's token-level prediction probabilities to estimate its confidence in classifying each paper as a conceptual advance. This yielded a continuous variable, representing the likelihood that a paper introduces a conceptual rather than methodological innovation.

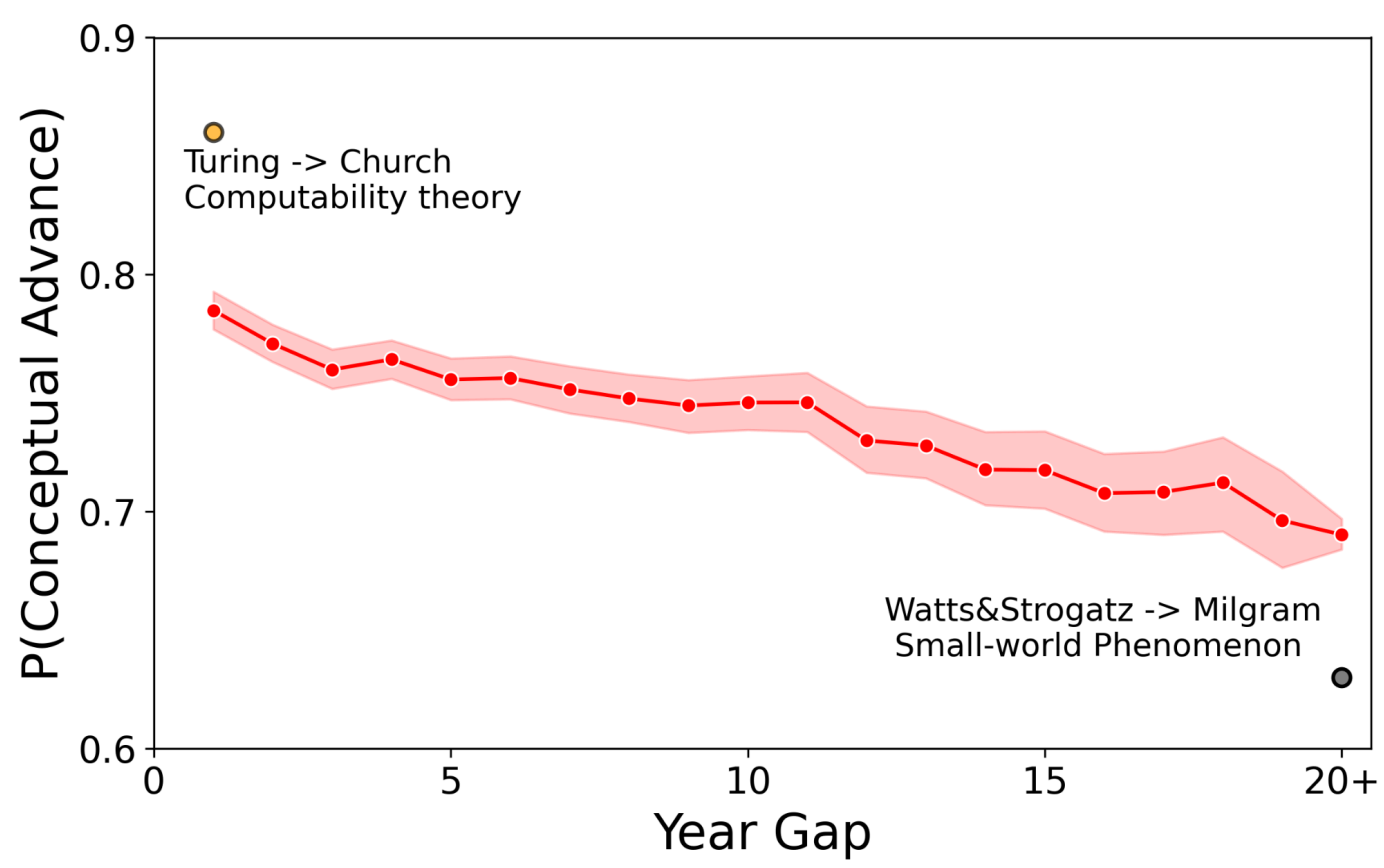

**Figure 4. Conceptual advances happen early; formalism advances emerge later.** Using Llama-3.1-70B-Instruct, we classified 52,488 highly cited papers (1900–2020) relative to their most cited reference. The model assessed whether each paper represented a conceptual (theory) or methodological (formalism) advance and assigned a probability for theory innovation. The mean decreases as the publication gap between the focal paper and its reference increases. Shaded regions represent 95% confidence intervals.

As shown in Figure 4, the probability of conceptual innovation declines with the temporal distance between the focal paper and its reference. Conceptual advances occur shortly after their reference works, while methodological or technical refinements are more likely to emerge later. Examples include Turing's 1937 paper, which received a for conceptual innovation relative to Church's 1936 work, and Watts & Strogatz's 1998 formalization of the small-world phenomenon as a methodological refinement of Milgram's 1967 study.

To ensure robustness, we tested several alternative prompting strategies: (1) few-shot prompts using labeled examples (Turing–Church and Watts–Milgram); (2) rewording options as "conceptual difference" and "formalism difference"; and (3) adding a third option ("others"). All variations produced consistent results, confirming that the observed trend is robust to prompt design. Together, these results challenge the assumption that novelty drives breakthroughs. Instead, they point to a fundamental tension between recombination and displacement in scientific innovation.

## 5. Discussion

Our findings call into question a core assumption in much of the literature on scientific innovation: that novel recombinations of ideas are inherently conducive to scientific breakthroughs. While recombination remains a valuable mechanism for generating new perspectives, our analysis suggests that it is often at odds with the mechanism responsible for displacing entrenched knowledge.